\documentclass[12pt,preprint,tighten]{aastex}
\usepackage[]{times, graphicx}
\newcommand{\soho}{{\em SOHO{}}}
\newcommand{\MWSO}{MWO{}}

\newcommand{\stereo}{{\em STEREO{}}}

\newcommand{\pref}{\protect\ref}

\begin{document}

\shorttitle{The Evolving Scale of The Quiet Solar Network}
\shortauthors{McIntosh et al.}
\title{Observing Evolution in the Supergranular Network Length Scale During Periods of Low Solar Activity}

\author{Scott W. McIntosh\altaffilmark{1}, Robert J. Leamon\altaffilmark{2}, Rachel A. Hock\altaffilmark{3}, Mark P. Rast\altaffilmark{3, 1}, Roger K. Ulrich\altaffilmark{4}} 
\altaffiltext{1}{High Altitude Observatory, National Center for Atmospheric Research, P.O. Box 3000, Boulder, CO 80307}
\altaffiltext{2}{Department of Physics, Montana State University, Bozeman, MT 59717}
\altaffiltext{3}{Laboratory for Atmospheric and Space Physics, Department of Astrophysical and Planetary Sciences, University of Colorado, Boulder, CO 80309}
\altaffiltext{4}{Division of Astronomy and Astrophysics, University of California, Los Angeles, CA 90095}

\begin{abstract}
We present the initial results of an observational study into the variation of the dominant length-scale of quiet solar emission: supergranulation. The distribution of magnetic elements in the lanes that from the network affects, and reflects, the radiative energy in the plasma of the upper solar chromosphere and transition region at the magnetic network boundaries forming as a result of the relentless interaction of magnetic fields and convective motions of the SunÕs interior. We demonstrate that a net difference of $\sim$0.5Mm in the supergranular emission length-scale occurs when comparing observations cycle 22/23 and cycle 23/24 minima. This variation in scale is reproduced in the datasets of multiple space- and ground-based instruments and using different diagnostic measures. By means of extension, we consider the variation of the supergranular length-scale over multiple solar minima by analyzing a subset of the Mt Wilson Solar Observatory (\MWSO) \ion{Ca}{2} K image record. The observations and analysis presented provide a tantalizing look at solar activity in the absence of large-scale flux emergence, offering insight into times of ``extreme'' solar minimum and general behavior such as the phasing and cross-dependence of different components of the spectral irradiance. Given that the modulation of the supergranular scale imprints itself in variations of the SunÕs spectral irradiance, as well as in the mass and energy transport into the entire outer atmosphere, this preliminary investigation is an important step in understanding the impact of the quiet sun on the heliospheric system.
\end{abstract}


\section{Introduction}
The spectral constituents of the total solar irradiance (TSI) can vary considerably over a solar cycle \citep[e.g.,][]{Frohlich2004}. As Figure~1 demonstrates, the \soho/VIRGO \citep[][]{1995SoPh..162..101F} TSI and Penticton 10.7cm solar radio flux \citep[$F_{10.7}$;][]{1969JRASC..63..125C} change considerably over solar cycle 23 and into~24 (1996 into 2010). They are globally in phase with the sunspot number (SSN), but show subtle changes in response during the lengthy 23/24 solar minimum. That difference is likely the result of the physical origins of the radiation being monitored as a proxy to solar activity, an issue of current debate (and concern) in the Sun-Climate community \citep[e.g.,][]{2010Natur.467..696H}. In addition, there is contention as to whether or not the radiative output of the Sun varied across minima \citep[][]{Frohlich2009} or if instrumental bias was at play - a debate that has grown out of the necessary cross-calibration of the discontinuous space record \citep[][]{2009GeoRL..3605701S}. It seems that the recent solar minimum has offered a chance to ``calibrate'' the discussion.


At times of high solar activity, the X-ray and extreme-ultraviolet (EUV) radiance (driven by changes in large-scale coronal magnetic morphology) dominate the TSI variation while, during quieter epochs, the TSI variation is dominated in the ultraviolet (UV) and infra-red (IR) radiance. The UV and IR components are, themselves, driven by contributions from the ever-present, relentlessly evolving, supergranular and granular length scales of emission in the transition region, chromosphere and photosphere \citep[e.g.,][]{Frohlich2004}. At times of solar minimum, it is the interplay of these components (with radiative losses that are sensitive function of scale and magnetic flux distribution) that dominate the variance of the TSI and becoming the radiation bands that directly impact the chemistry of the EarthÕs upper atmosphere \citep[e.g.,][]{2007JGRD..11223306M}.


Unfortunately, a complete underlying physical understanding of the process(es), that are driving the radiative variation of our star is distant. As our knowledge of the mass and energy transport through the solar atmosphere improves, we will gain a better understanding of the intricate physics behind each component of the spectral irradiance. In this Letter we investigate the dominant length scale of emission in the quiet chromosphere and transition region, the ubiquitous magnetic network \-- the fundamental length scale of the mass and energy release into the quiet corona and solar wind \citep[e.g.,][]{McIntosh2007, McIntosh2009, 2011Sci...331...55D, McIntosh2011}.

In the following sections we discuss the observations studied and the methods applied to them, before discussing our results and placing them in context with a preliminary analysis of a much longer image timeseries, the Mt Wilson Solar Observatory (\MWSO) \ion{Ca}{2}~K spectroheliogram archive.

\section{Observations}
We make use of data from five independent sources to study the variation of the emission and magnetic scales present in the transition region and photosphere over the course of solar cycle 23 and into the early portion of cycle 24. Our primary data source is the \soho{} spacecraft \citep[][]{Fleck1995} with particular emphasis on the 96-minute MDI \citep[][]{Scherrer1995} line-of-sight (LOS - 300s exposure) magnetograms and EIT \citep[][]{Boudine1995} \ion{He}{2} 304\AA{} (four per day) synoptic data sets. We supplement these observations in the declining phase of cycle 23 with the \stereo/SECCHI EUVI \citep[][]{secchi} \ion{He}{2} 304\AA{} image sequences along with \ion{Ca}{2} K 3936\AA{} channel images from the Precision Solar Photometric Telescope \citep[PSPT;][]{2008ApJ...673.1209R} at the Mauna Loa Solar Observatory (MLSO).


\section{Method}
For a \soho{} 304\AA{} image and LOS magnetogram pair taken near 13UT on April 10 2008, we compute the mean supergranular cell radius ($\langle r \rangle$), using watershed segmentation of the image,  and ``Magnetic Range of Influence'' \citep[MRoI;][]{McIntosh2006} techniques. To minimize the effects of LOS foreshortening of the cells and contamination of the magnetograms we will only consider the statistics of the region within 60\% of a solar radius ($R_s$) from disk center.

As originally presented with application to solar UV images, watershed segmentation \citep[e.g.,][]{Lin2003,Lin2005} uses an intensity image as a topographic map where the brightest regions in the image are peaks and the darkest are troughs \citep{McIntosh2007}. The algorithm then drops``water'' onto the map at random locations, that water then ``flows'' downhill until it finds the nearest trough, which then starts to fill. When neighboring troughs fill, their intersection forms a watershed, mapping out the topographic boundary between them, outlining the network. 

Applied to a solar image (see Fig.~\pref{f2} panel A) out to 0.98$R_s$ these watershed boundaries outline the supergranular network (panel B). We calculate the cell area and radius of each watershed basin by counting the number of pixels in the basin interior and computing the radius of gyration for those pixels (panel C). The value of $\langle r \rangle$ used below is derived from the mean of the resulting distribution (panel D) in the central 0.6 $R_s$. The gray shaded outline of the distribution shows the distribution range after performing a Monte-Carlo test - computing 1,000 segmentations of the image where, in each instance, 10,000 randomly chosen pixels have had Poisson noise added. For the image shown, the $\langle r \rangle$ = 24.08Mm with a variance of 0.01Mm. Note that the variance of the mean in any one image is considerably smaller than the annual and monthly variations in the value itself. For PSPT images we compute $\langle r \rangle$ using the ``iterative medial axis transform'' of \citet{1999A&A...344..965B, 2005SoPh..228...81B}, which is completely independent of the watershed segmentation technique based instead on an iterative estimate of  pixel connectivity. For the sake of brevity the interested reader is referred to \citet{2009ApJ...707...67G} for further detail on the analysis technique as it is applied to PSPT images.


An alternate measure of spatial scale in the lower solar atmosphere is the MRoI (Fig.~\pref{f3}). The MRoI is derived from LOS magnetograms (panel A) and is defined as the distance required by the magnetic field measured in one pixel to find the magnetic flux of equal magnitude and opposite sign such that the sum over that distance is zero. Therefore, the MRoI (panel B) estimates the distance needed to ``balance'' or ``close'' the magnetic field, providing information about the separation of magnetic elements in the photosphere and, therefore, of the vertices of the supergranular network. For the disk-center region of the magnetograph shown here the distribution of MRoI values in the region (panel C) is well represented by a power-law with index -1.718 ($\pm$ 0.094) where the error is that of the linear fit to the log-log distribution.

\section{Analysis}
Several thousand \soho/EIT 304\AA{} synoptic images taken over the last fourteen years (and the twin \stereo{} spacecraft images over the past four) are analyzed, including the ``unusual'' solar activity minimum of 2009. Applying the watershed segmentation and MRoI processing techniques (discussed above) on these large image datasets we are able to characterize variations in $\langle r \rangle$ and a measure of the length-scale distribution of the magnetic elements that comprise the vertices of the supergranule cells. The observations of the \stereo{} spacecraft (in the same \ion{He}{2} 304\AA{} channel) and ground-based observations from PSPT instrument are used to validate the observed variance, particularly through the 2009 solar minimum.

\begin{figure}
\plotone{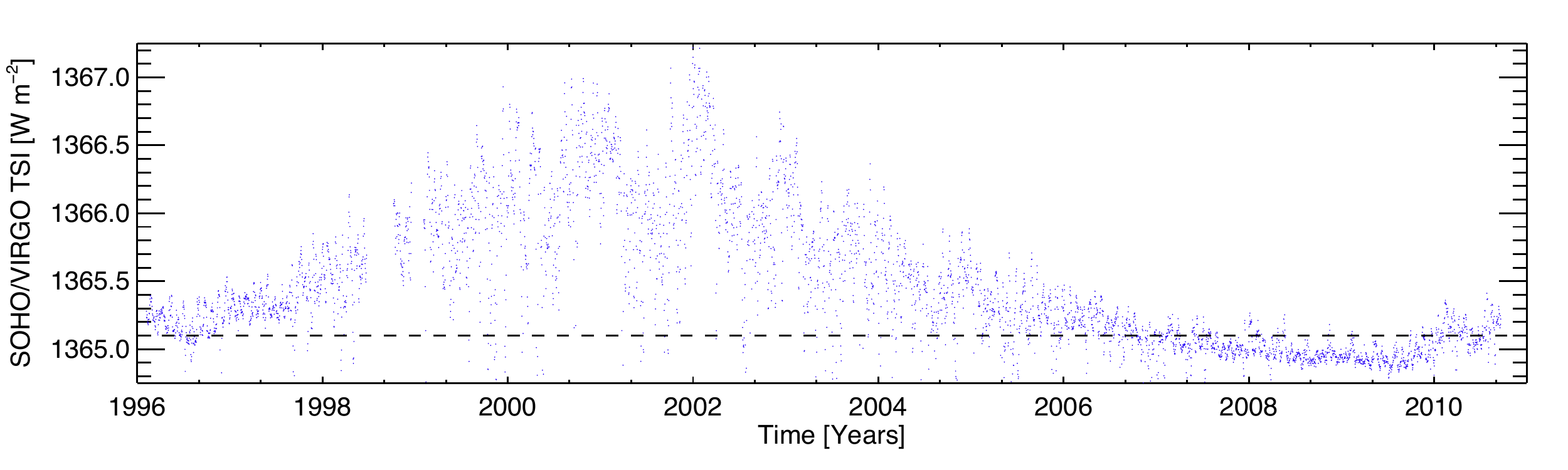}
\plotone{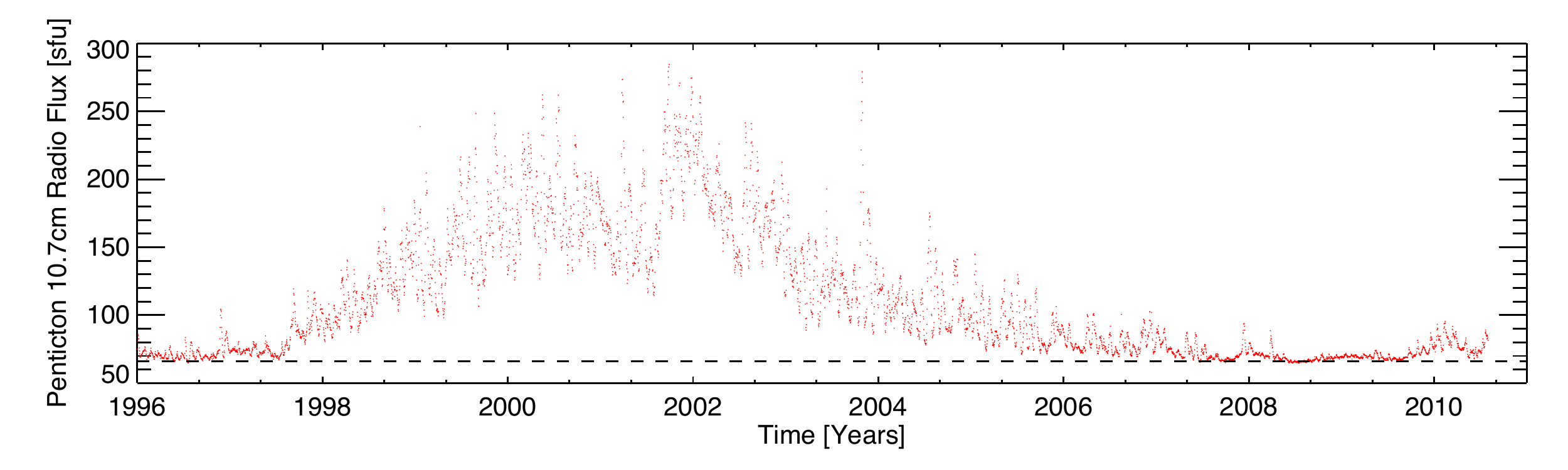}
\plotone{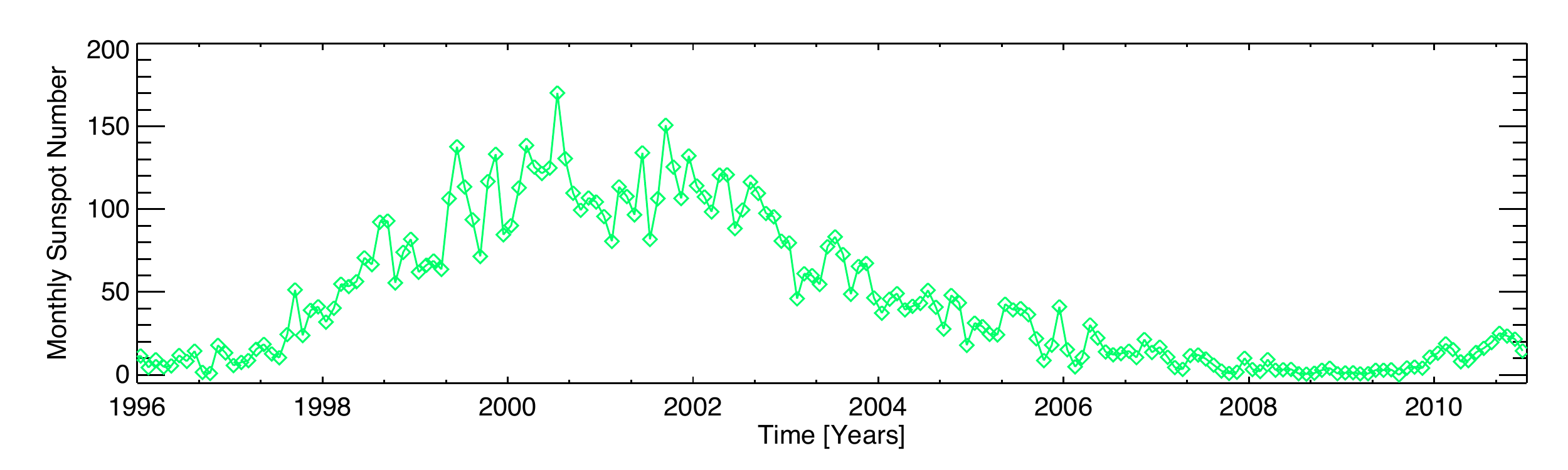}
\caption{From bottom to top these plots show the variation in the mSSN (green), $F_{10.7}$ solar radio flux (red), and the \soho/VIRGO TSI (blue) over solar cycle 23. Horizontal dashed lines illustrate typical values at the deepest portion of the cycle 22/23 solar minimum; 1365.1~Wm$^{-2}$, and 65 sfu. While the mSSN and $F_{10.7}$ appear to ``bottom-out'' in late 2007 through late 2009 the TSI continued decreasing, bottoming out in the fall of 2009. \label{f1}}
\end{figure}

\begin{figure}
\plotone{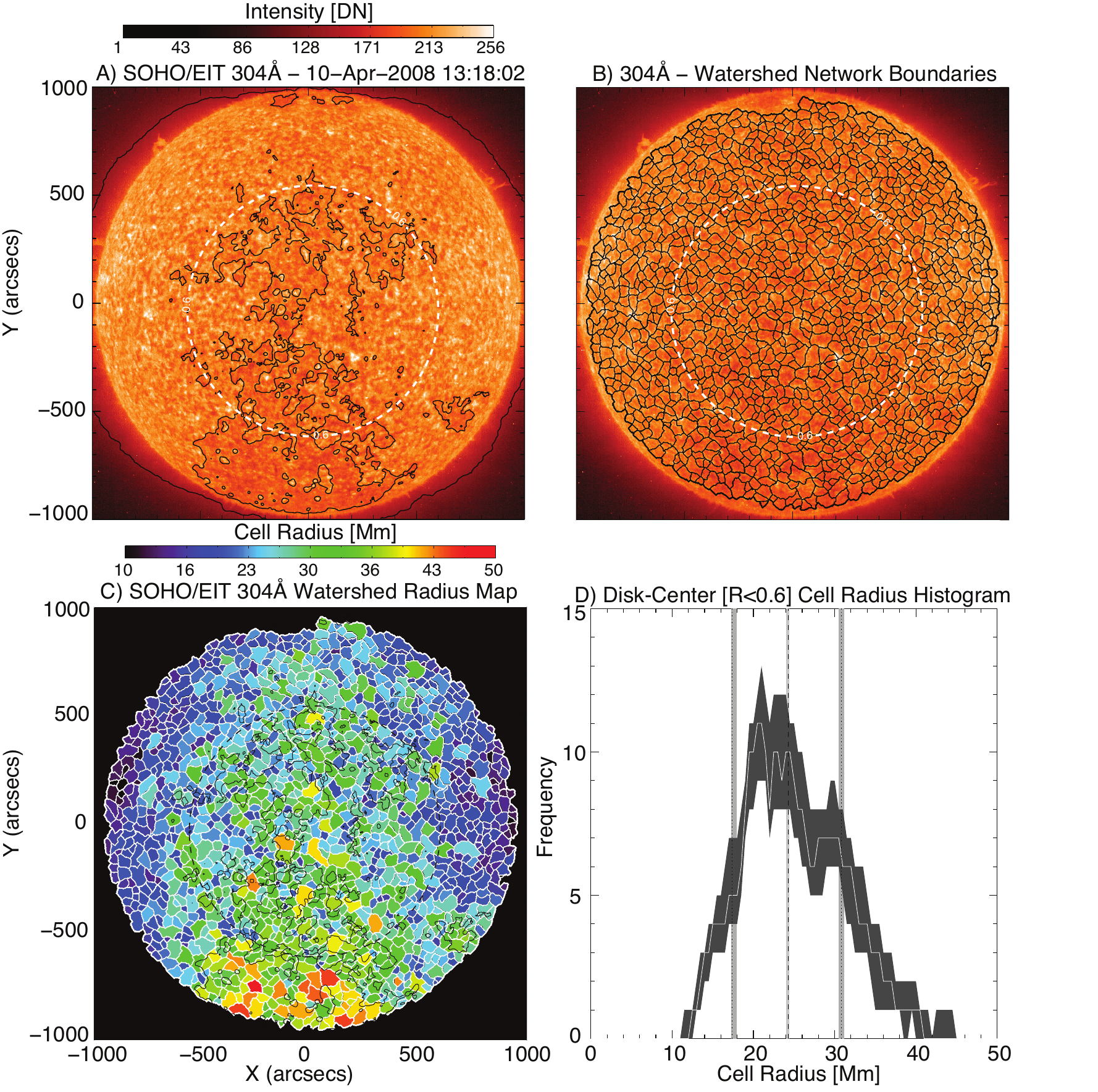}
\caption{Watershed segmentation applied to a \soho/EIT 304\AA{} image taken on 2008 April 10, 13:18UT. Panel A shows the 304\AA{} image overlaid with the (black) EIT 195\AA{} 150~DN iso-intensity contour to illustrate weak, possibly coronal hole, emission and the white dashed circle of radius $0.6 R_s$. Panel B shows the same image following application of the watershed segmentation; the boundaries of the supergranular cells are black. Panel C shows the cells color-coded by radius. Panel D shows the distribution of cell radii in the central portion of the image and how the envelope, width, and mean of the distribution change following 1000 Monte Carlo realizations of the image. \label{f2}}
\end{figure}

\begin{figure}
\plotone{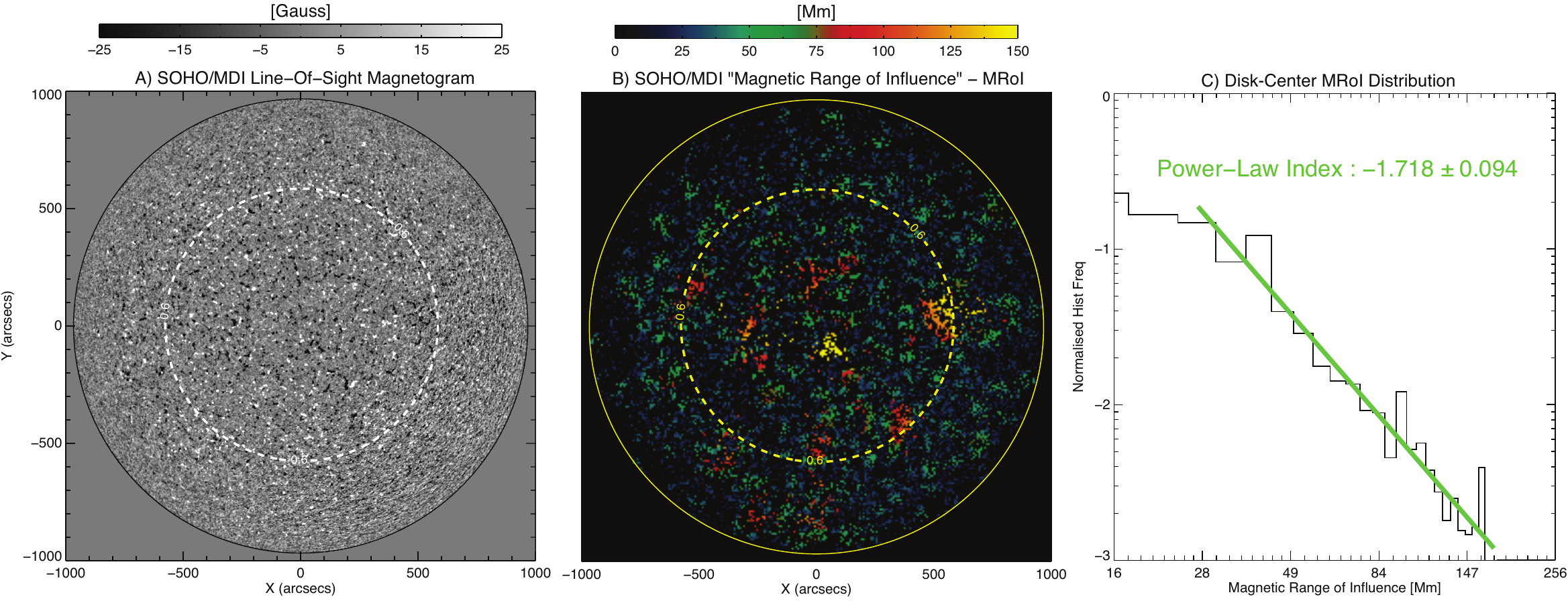}
\caption{MRoI applied to a \soho/MDI LOS magnetogram taken closest to the image used in Fig.~\pref{f2} (April 10 2008, 12:51UT). Panel A shows the MDI LOS magnetogram used while panel B shows the MRoI map for that magnetogram. Again we show the dashed circle of radius 0.6 $R_s$. Using the disk center values we determine the distribution of MRoI values (panel C) along with a linear fit (green line). \label{f3}}
\end{figure}

Figure~\pref{f4} illustrates the variation in these various scales over solar cycle 23 and into the start of cycle 24. The top four panels, from top to bottom, show the variation in $\langle r \rangle$ for \soho/EIT (black dots), {\em STEREO-A} (green dots), {\em STEREO-B} (blue dots), PSPT (red dots). In addition, these panels show the variation in the solar radius as observed from each platform (dot-dashed line) that are used to correct for subtle annual variation in the size of the Sun in the images that is not taken into account in the segmentation analysis. The variation in solar radius is used to correct the raw segmentation values in a simple fashion \-- in each case $\langle r \rangle$ and solar radius form a linear relationship, the gradient of this relationship is then used to adjust the measured timeseries. The most notable excursion of $\langle r \rangle$ is for {\em STEREO-B} where the solar radius varies annually by nearly 9\%. For the PSPT analysis, shown in the fourth row, we need to multiply $\langle r \rangle$ by an factor of 1.9 to match the scale determined from the space-based observations  likely due to photospheric contributions to the measurements and the differing segmentation algorithm employed. We note that in this analysis is focused on the difference between solar minima and the trends which Fig.~\pref{f4}clearly shows as common to all the measurements. It is unlikely that errors are common to all the data sets or they are systematic in nature. The fifth row of the figure shows the correspondence between the four adjusted timeseries in comparison with the \soho/VIRGO TSI (yellow; cf. Fig.~\pref{f1}). 

The correspondence of the diagnostics, their minimum scale, phase, and variance going into the recent solar minimum is striking, and we feel that this is a strong indication that the reduction of $\langle r \rangle$ of the recent solar minimum, by about 0.5Mm on average, is a result of changes on the Sun and not instrumental in nature. We also note the correspondence between the period where $\langle r \rangle$ falls below 25Mm is exactly when the \soho/VIRGO TSI is reduced below 1365.1Wm$^{-2}$ and both recover in phase with one another in the last quarter of 2009. This suggests that the reduction in the dominant emission scale of the quiet sun network is directly responsible for that portion of the TSI variation. 


Finally, we investigate the magnetic roots of the supergranular structure itself through the MRoI (bottom row) \- making the assumption that the power-law index ($\delta$) of the MRoI distribution provides information about the separation between magnetic flux aggregations in the MDI magnetograms, where a flat distribution (small $\delta$) would indicate significant spread in the scale, and steep distribution (larger $\delta$) would indicate that there are more shorter length scales present in the MRoI maps and that the net separation of the flux elements is smaller. From the plotted distribution (black triangles) we see that $\delta$ increases from -1.6 in the cycle 22/23 minimum, flattening out at solar maximum, before approaching -1.95 in the middle of 2009. From the last quarter of 2007 through the last quarter of 2009 we see that $\delta$ is lower than the values of the previous minimum \-- we interpret this as a very strong indication that the magnetic field decayed to smaller length scales than those of the previous minimum, a deduction that validates our previous diagnosis that $\langle r \rangle$ is reduced substantially.

\subsection{\MWSO{} Calcium Archive Network Variation}
A natural, but preliminary, expansion of our investigation lies in the analysis of the Mount Wilson Solar Observatory (\MWSO) \ion{Ca}{2}K spectroheliogram digital archive.\footnote{Information about the \MWSO{} \ion{Ca}{2}K image archive can be found \url[http://ulrich.astro.ucla.edu/MW_SPADP/index.html]{online}.} Applying watershed segmentation to the digitized images covering three complete solar cycles (1944 to 1976) we see the results in Fig.~\pref{f5} (red dots) plotted versus the monthly SSN (mSSN; green dots) over the same epoch. There is significantly more spread in $\langle r \rangle$ determined from these images than those shown above. The most pronounced signature in the timeseries is the slower decay of $\langle r \rangle$ in the descending phase of the three cycles hinting that it provides some measure of the global diffusion timescale of the magnetic fields \citep[][]{1997ApJ...481..988H, 1999SoPh..187....1S, 2010ApJ...725.1082H}. This feature is also evident in the contemporary data (Fig.~\pref{f4}) prompting a study of coronal holes and the impact of local unipolarity on the supergranule cells, where they appear to be bigger (Fig.~\pref{f1}C). We must be careful in not drawing too strong a conclusion from the watershed analysis of the \MWSO{} archive. It does, however, provide significant motivation for a future study involving other digital archives that span a similar timeframe, like those from Arcetri and Kodaikanal \citep[e.g.,][]{2009ApJ...698.1000E, 2009SoPh..255..229F}, linking them, through observations taken by the National Solar Observatory \citep[e.g.,][]{1998ApJ...496..998W}, through the 1980s to overlap with PSPT and \soho{}. To validate any impact of coronal holes on the historical data record, digitization projects should be expanded to H$\alpha$ where the lack of fibril structure in the broadband images is used to identify coronal holes \citep[][]{1998SoPh..177..375F}.


\section{Discussion} 
Supergranulation is a flow pattern visible at the solar surface that is either driven by the convective motions of the solar interior, or emerging directly from the self-organization of granular flows \citep[e.g.,][]{2000A&A...357.1063R, 2003ApJ...597.1200R, 2007ApJ...662..715C}. The network pattern that results from the flow field is the dominant scale of the ``quiet'' solar atmosphere, forming a network of magnetic conduits through which mass and energy are driven into the corona and heliosphere. A direct consequence of the resulting circulation of (heated and cooling) plasma in the outer solar atmosphere, is the production of the UV/EUV radiation that continuously bathes the Earth. Therefore, monitoring variations in this length-scale is an important factor in assessing the SunÕs direct (radiative in the form of the TSI), and indirect (from the solar wind and the components of the spectral irradiance) forcing of the Earth and its climate. In the case of the recent solar minimum we believe that the changes in the chromospheric network, and possible weaker quiet-Sun fields, are largely responsible for the continued reduction of the TSI over what might have been expected based on previous solar minima.

From the variation in length-scale observed we deduce that the magnetic roots (vertices) of the supergranular network decayed to a smaller length-scale than in the previous minimum and likely relating to the low activity of the recent solar minimum \citep[see e.g.,][among many others]{2008GeoRL..3518103M,2009JGRA..11409105G, 2010GeoRL..3716103S, 2010GeoRL..3718101M}. We stress that while the length-scale derived from our analysis may not precisely represent that scale on the Sun, the systematic nature of our study suggests that the differential measurement is robust. Further, the continuation of the magnetic field diffusion to smaller length scales may in part explain the upturn in the longer wavelength (optical) radiation observed as the shorter (UV/EUV) wavelength emissions decrease \citep[][]{2010Natur.467..696H}. If the very small magnetic elements diffuse away from the supergranule vertices, becoming more evenly spread, then the ``hot wall'' effect on those discrete bundles of magnetic field \citep[][]{1976SoPh...50..269S} can, in principal, increase the net long-wavelength radiation of the ensemble of individuals compared to the original network conglomerate. It is unclear what physical process, or processes, in the solar interior drive the occurrence, or modulation, of the supergranular network. However, recent results \citep[][]{2010Sci...327.1350H,2010ApJ...725.1082H} have pointed to subtle variations in the SunÕs internal convection pattern that directly affect the creation, movement, and dispersal of magnetism over the course of cycles and, with particular effect over this ``unusual'' solar minimum.

The energetic ramifications of this apparent evolution in supergranular network scale on the heliospheric system: a less dense solar wind and corona \citep[][]{2008GeoRL..3518103M}, significantly reduced soft X-Ray, EUV, and UV emission \citep[][]{2010GeoRL..3716103S}, etc might be consistently explained if we think more about how a supergranule is constructed. What is the fundamental mechanism is for the heating and release of magnetic energy in that structure \citep{2011Sci...331...55D}, and how is that energy distributed in the outer atmosphere \citep[][]{McIntosh2007, McIntosh2011}? The apparent paradox of smaller scale equating to less emission, mass input, etc. leads us to speculate that the energetic output of supergranules is not as simple as the UV/EUV radiation and mass lost into the solar wind scaling linearly with a change in the field strength, rather depending more on the complexity of the smaller scale magnetic flux elements that comprise the network vertices. For the case of the recent solar minimum it is possibly that we crossed a threshold of energy production driven, in large part, by the fact that the supergranular vertices diffused, reducing the frequency and strength of the heating events rooted there. This is a topic to be explored in future work, where it is hoped that the Interface Region Imaging Spectrograph (\url[http://iris.lmsal.com/]{IRIS}) will permit an investigation into the scaling of energy release into the outer atmosphere with the complexity of the magnetic field in the photosphere and chromosphere. 

\begin{figure}
\epsscale{0.75}
\plotone{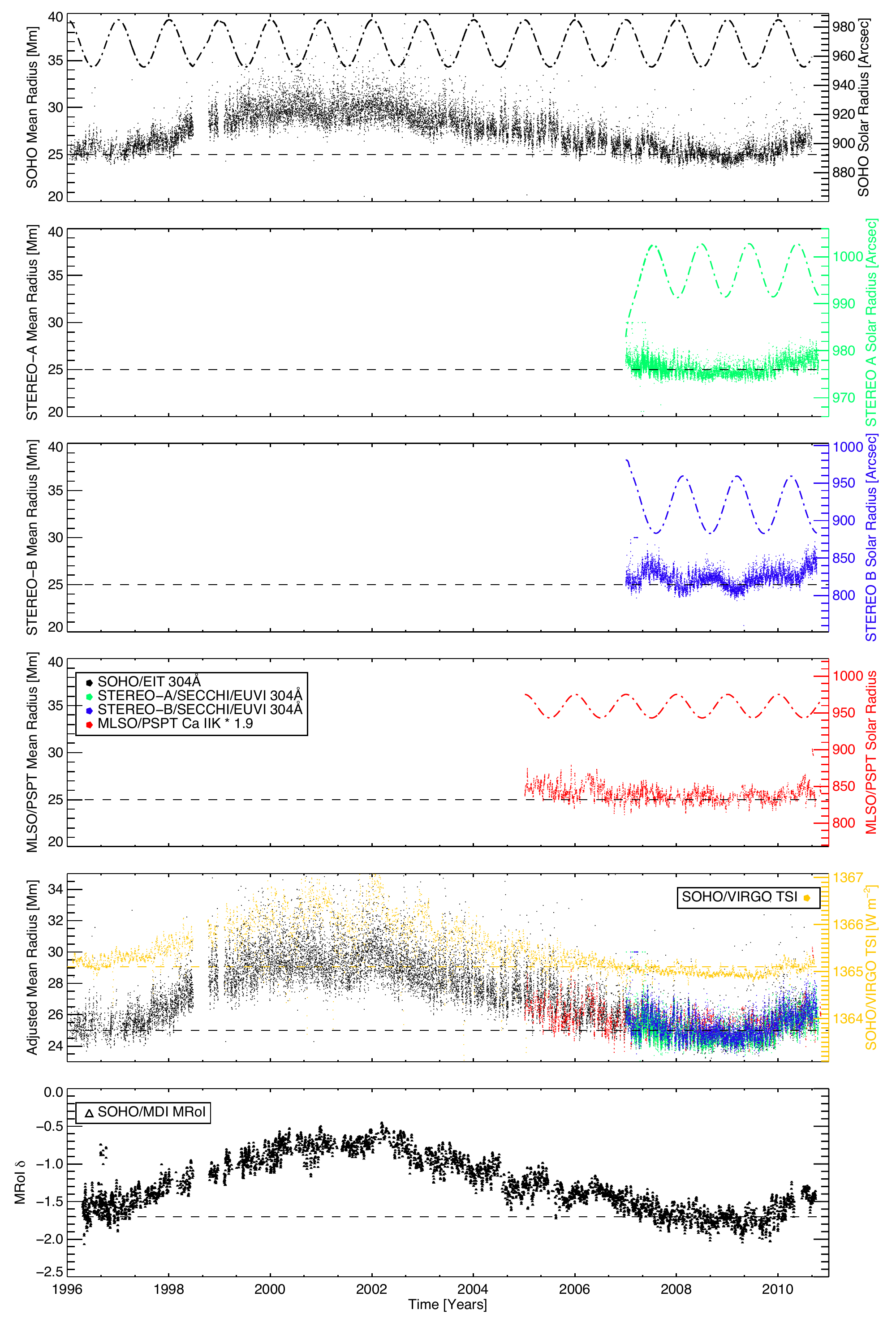}
\caption{From top to bottom, the variation in $\langle r \rangle$ determined from the \soho/EIT, \stereo{} A EUVI, \stereo{} B EUVI, PSPT, image sequences and the power-law exponent ($\delta$) of the \soho/MDI MRoI. The upper four panels show the uncorrected $\langle r \rangle$ and variation in the solar radius (dot\--dashed line) as seen from each observing platform used to correct each timeseries. In the fifth row, we show the adjusted timeseries and compared to the \soho/VIRGO TSI timeseries (orange dots) from Fig.~\pref{f1}. For reference we draw dashed lines for a $\langle r \rangle$ = 25Mm, TSI of 1365.1Wm$^{-2}$, and MRoI $\delta$ of -1.7 on the appropriate panels. \label{f4}}
\end{figure}

\begin{figure}
\epsscale{1.}
\plotone{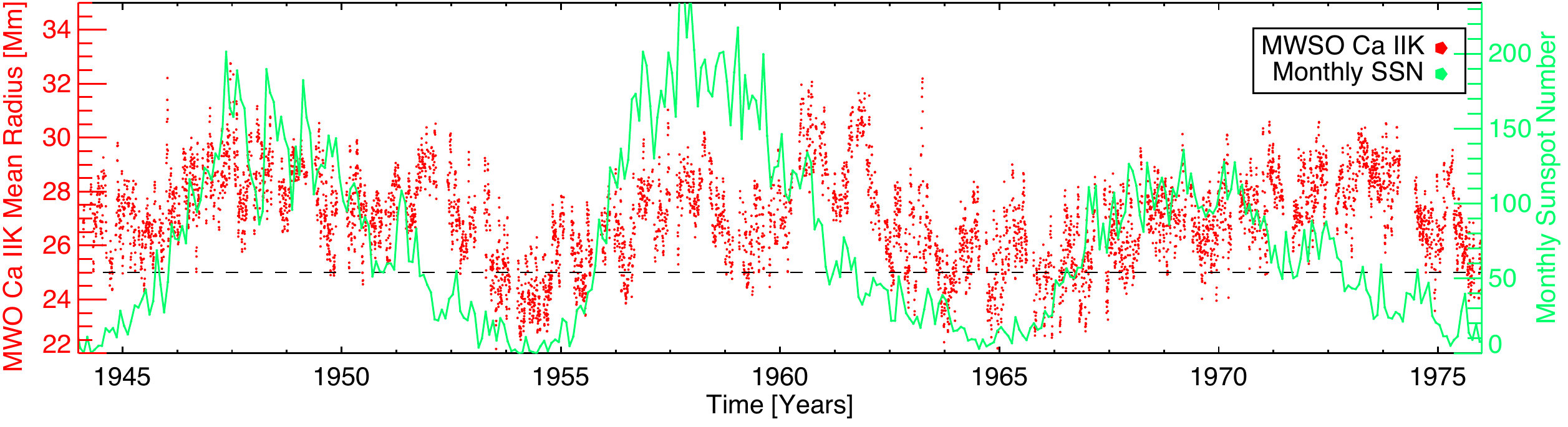}
\caption{Variation in $\langle r \rangle$ determined from a watershed segmentation of the \MWSO{} \ion{Ca}{2}K image archive (red dots) covering three solar cycles from 1944 to 1976 with respect to the corresponding variation in the mSSN (green dots). Again, for reference we draw a dashed horizontal line for $\langle r \rangle$ = 25Mm. \label{f5}}
\end{figure}

It is similarly enticing to speculate about the connection of this result to the Maunder Minimum as a possible cause of the Little Ice Age \citep[][]{1976Sci...192.1189E}. Clearly, we experienced a significant number of spot-free days from 2008 and through 2009. The evidence presented in this Letter is consistent with a picture where the decay of magnetic flux to smaller and smaller length-scales because larger scale organized flux is not injected into the system. Note that the system responds dramatically once that injection takes place, as can be readily observed in Fig.~\pref{f4}. What happens to these scales when organized large-scale flux is not erupting through the photosphere for several years, or decades, catastrophic breakdown in mass transport in the outer atmosphere with a resulting knock-on in radiative and particulate output? Recent estimates have been made to recreate the magnetic and radiative conditions during the Maunder Minimum \citep[see, e.g.,][]{2003ApJ...591.1248W, 2007SoPh..246..309T, 2010JGRA..11512112K}, but we must strive to understand the basic physical processes, and radiative scales, affecting the star's radiative and particulate output in contemporary observations or those of the near future (see above). Then, we can investigate how various components of the spectral irradiance have on the Earth's atmosphere \citep[][]{2007JGRD..11223306M, 2010Natur.467..696H} at all epochs of solar activity, from hyper-active maxima to grand minima.

%

\acknowledgements
Data from the VIRGO experiment was retrieved through PMOD/WRC, Davos, Switzerland. The effort was supported by external funds (SWM: NNX08AL22G, NNX08AU30G  from NASA and ATM-0925177 from the National Science Foundation; RJL: NNH08CC02C from NASA). NCAR is sponsored by the National Science Foundation. Processing of the \MWSO{} archive was funded by the National Science Foundation under Grant No.0236682 to RKU.

\clearpage

\end{document}